\documentclass[aps,prd,twocolumn,superscriptaddress,showpacs,showkeys]{revtex4-1}  
\usepackage{graphicx,color} 
\usepackage{xcolor}
\usepackage[hidelinks]{hyperref}
\usepackage{rotating}
\usepackage{amssymb}
\usepackage{amsfonts}
\usepackage{amsmath}
\usepackage{xspace}
\usepackage{mathtools} 
\usepackage{verbatim}

\DeclarePairedDelimiter\braket{\langle}{\rangle}
\DeclarePairedDelimiterX\Braket[2]{\langle}{\rangle}{#1 \delimsize\vert #2}

\usepackage{listings}
\makeindex

\usepackage[thinlines]{easytable}

\usepackage[super]{nth}

\newcommand\ie{\mbox{\textit{i.\,e.}}\xspace}
\newcommand\cf{\mbox{c.\,f.}\xspace}
\newcommand\eg{\mbox{e.\,g.}\xspace}

\newcommand\D{\mathrm{d}}

\newcommand{\hx}{\hat{x}}
\newcommand{\hp}{\hat{p}}
\newcommand{\hX}{\hat{X}}
\newcommand{\hP}{\hat{P}}
\newcommand{\mpar}{\dot{\partial}}

\begin{document}
	
	\title{Reinterpreting deformed Heisenberg algebras}
	
	
	
	
	\author{Fabian Wagner}
	\email[]{fabian.wagner@usz.edu.pl}
	\affiliation{Institute of Physics, University of Szczecin, Wielkopolska 15, 70-451 Szczecin, Poland}
	\date{\today}
	\begin{abstract}
		Minimal and maximal uncertainties of position measurements are widely considered possible hallmarks of low-energy quantum as well as classical gravity. While General Relativity describes interactions in terms of spatial curvature, its quantum analogue may also extend to the realm of curved momentum space as suggested, \eg in the context of Relative Locality in Deformed Special Relativity. Drawing on earlier work, we show in an entirely Born reciprocal, \ie position and momentum space covariant, way that the quadratic Generalized Extended Uncertainty principle can alternatively be described in terms of quantum dynamics on a general curved cotangent manifold. In the case of the Extended Uncertainty Principle the curvature tensor in position space is proportional to the noncommutativity of the momenta, while an analogous relation applies to the curvature tensor in momentum space and the noncommutativity of the coordinates for the Generalized Uncertainty Principle. In the process of deriving this map, the covariance of the approach constrains the admissible models to an interesting subclass of noncommutative geometries which has not been studied before. Furthermore, we reverse the approach to derive general anisotropically deformed uncertainty relations from general background geometries. As an example, this formalism is applied to (anti)-de Sitter spacetime.
	\end{abstract}
	
	
	\maketitle
	
\section{Introduction}
\label{sec:intro}

As the field of quantum gravity has been maturing, it has outlined a vast landscape of paths towards a final theory (see \cite{Kiefer04,Loll22} for recent broad reviews). However, as of yet none of the proposed approaches has provided decisively convincing evidence in its favour. The main reason for this continuing dissent lies in the absence of experimental input caused by the seeming inaccessibility of the relevant scale, the Planck length $l_p.$

In the wake of the continuously increasing precision in observations and experiments as well as control over quantum phenomena, though, the prospect of a meaningful phenomenology of quantum gravity appears ever more feasible \cite{Amelino-Camelia99} (see \cite{Addazi21,Amelino-Camelia08} for recent reviews). 
Among the manifold currents in the field, the minimal length paradigm figures as the oldest and continues to be of great importance. Over the time it has been motivated from String theory \cite{Amati87,Amati88,Gross87a,Gross87b,Konishi89}, Loop quantum gravity \cite{Maggiore93b,Hossain10,Majumder12a,Girelli12}, Asymptotic Safety \cite{Ferrero:2022hor}, noncommutative geometry \cite{Battisti08c,Pramanik13,Chaichian:2005yp} as well as Ho\v{r}ava-Lifshitz gravity \cite{Myung09b,Myung09c,Eune10}, but also derived from general arguments combining gravity and quantum theory \cite{Mead64,Mead66,Padmanabhan87,Ng93,Maggiore93a,Amelino-Camelia94,Garay94,Adler99a,Scardigli99,Capozziello99,Camacho02,Calmet04}. Indeed, it is underlying the aforementioned LIV \cite{Colladay98} as well as Deformed Special Relativity (DSR) \cite{Amelino-Camelia00,Magueijo01} (see \cite{Arzano21,Bevilacqua22} for recent reviews) and within single particle quantum mechanics the Generalized Uncertainty Principle (GUP) \cite{Maggiore93c,Kempf94,Kempf96a,Benczik02,Das12,Buoninfante20,Petruzziello20,Bosso20,Bosso21} (see \cite{Hossenfelder12} for a review). 

The latter effect which is investigated the present paper posits the minimal length as an epistemological obstruction to precise measurements, not as a fundamental discreteness \cite{Hagar14}, by modifying Heisenberg's famous relation as (here understood perturbatively in one dimension)
\begin{equation}
	\Delta x\Delta p\geq \frac{\hbar}{2}\left[1+\left(\frac{l_\text{UV}\Delta p}{\hbar}\right)^2\right],
\end{equation}
with the minimal length $l_{\text{UV}},$ supposedly related to the Planck scale, and the fundamental uncertainties of position and momentum measurements $\Delta x$ and $\Delta p,$ respectively. 

However, the quantum mechanical formalism (as well as the Hamiltonian one) enjoys a symmetry \cite{Born38} with respect to the exchange of positions and momenta ($\hx\leftrightarrow -\hp$), dubbed Born reciprocity, whose preservation implies that a correction of this kind must be complemented by a position-dependent counterpart, resulting in the Generalized Extended Uncertainty principle (GEUP) \cite{Bambi08a,Mignemi09,Ghosh09,CostaFilho16}
\begin{equation}
	\Delta x\Delta p\geq \frac{\hbar}{2}\left[1+\left(\frac{l_\text{UV}\Delta p}{\hbar}\right)^2-\left(\frac{\Delta x}{l_\text{IR}}\right)^2\right]\label{GEUP1D},
\end{equation}
with an additional maximal length $l_\text{IR}.$ As a matter of fact, the mixing of IR and UV effects, \ie such acting on large and short distance scales, constitutes another widely suspected feature of quantum gravity \cite{Minwalla99} -- and indeed, the cosmological horizon, predicted by General Relativity, can be understood as a limit to distance measurements akin to a maximal length. 

The position-dependent corrections leading to so called Extended Uncertainty Principles (EUPs) were shown to be related to spatial curvature in a recent series of papers \cite{Schuermann18,Dabrowski19,Dabrowski20,Petruzziello21,Wagner21b}. While those works do not employ the usual uncertainty relations and alternative ways of introducing an inequality of the type \eqref{GEUP1D} exist \cite{Lake18,Lake19,Lake20}, the most common approach also pursued in this article consists in modifying the underlying algebra of observables. 

In this context, the present author has shown in an earlier piece \cite{Wagner21a} (see also \cite{Wagner22}) that theories of single particles obeying a GUP in flat space can be mapped onto the dynamics of an ordinary particle on nontrivial momentum space, albeit with a somewhat idiosyncratic, non-momentum space-covariant kinetic Ha\-mil\-tonian which simplified the general treatment. For the quadratic GUP, the resulting curvature in momentum space is proportional to the noncommutativity of the coordinates in the original theory. Similar results have also been obtained in \cite{Singh21,Gubitosi21}. Apart from that, the concept of curved momentum spaces has been among the hot topics of quantum gravity phenomenology since the field's infancy \cite{Kowalski-Glikman02b,Kowalski-Glikman02c,Amelino-Camelia11,Pfeifer21,Freidel13,Freidel14,Freidel15,Carmona19,Relancio20a,Relancio21a,Relancio21b,Carmona21,Relancio22,Franchino-Vinas22}.

In the present paper, we modify this analysis such that the resulting Hamiltonian in the dual theory is covariant in position as well as momentum space, by assuming the kinetic part of the Hamiltonian to be proportional to the squared geodesic distance to the origin in momentum space as suggested in \cite{Pfeifer21,Relancio21a,Relancio22}. Correspondingly, in a fully reciprocal theory, the potential of the isotropic harmonic oscillator has to be proportional to the squared geodesic distance to the origin in position space, which serves as an additional constraint. Further treating both positions and momenta on exactly reciprocal footing during the entire analysis and working perturbatively with the quadratic GEUP, we obtain an map from the GEUP-deformed theory to the quantum dynamics of an ordinary particle on a nontrivial cotangent manifold, analogous to the less general result of \cite{Wagner21a,Wagner22}.  The covariance of the approach, however, restricts the admissible GEUP-models to an interesting subclass, whose GUP-sector, for example, apparently does not feature a minimal length but noncommutative coordinates.

Furthermore, in the EUP-limit (vanishing minimal length), the metric is solely position-dependent and the Riemann tensor results proportional to the noncommutativity of the momenta in the original theory, while the same occurs to the curvature tensor in momentum space and the noncommutativity of the coordinates in the GUP-limit (infinite maximal length).  This fact is, again, in line with \cite{Wagner21a,Wagner22}, underscoring the robustness of the result.

Finally, we reverse the formalism, thus deriving general anisotropic GEUP-algebras (see \cite{Gomes22a,Gomes22b}) in terms of the background curvature tensors. This result is applied to (anti-)de Sitter geometries at the relativistic level. There is an interesting connection of these findings to momentum gauge fields \cite{Guendelman:2022gue} which is studied further in \cite{Guendelman:2022ruu}. 

The paper is organized as follows: First, we introduce deformed Heisenberg algebras and explain how to work with them in Sec. \ref{sec:GUPQM}. We further use Sec \ref{sec:reint} to outline in rather general terms how the mentioned map is to be implemented. Sec. \ref{sec:qGEUP} is devoted to the perturbative application to the quadratic GEUP. The formalism is applied inversely in Sec. \ref{sec:revGEUPs}. Finally, we wrap up and discuss the findings in Sec. \ref{sec:Discussion}.

\section{Deformed Heisenberg algebras\label{sec:GUPQM}}

Instead of the common case known from textbook quantum mechanics, we assume that the physical position and momentum operators satisfy the deformed algebra
\begin{subequations}
	\begin{align}
		[\hx^i,\hx^j]=&i\hbar\theta^{ij}(\hx,\hp),\\
		[\hp_i,\hp_j]=&i\hbar\tilde{\theta}^{ij}(\hx,\hp),\\
		[\hx^i,\hp_j]=&i\hbar f^i_j(\hx,\hp),\label{GUPalg}
	\end{align}
\end{subequations}
with the sufficiently well-behaved functions $\theta^{ij},$ $\tilde{\theta}_{ij}$ and $f^i_j,$ for which the Jacobi identities imply that 
\begin{subequations}
	\label{JacId}
	\begin{align}
		[\hx^{[i},\theta^{jk]}]=&0,\\
		[\hp_{[i},\tilde{\theta}_{jk]}]=&0,\\
		[\theta^{ij},\hp_k]=&2[f^{[i}_k,\hx^{j]}],\\
		[\tilde{\theta}_{ij},\hx^k]=&2[f^{k}_{[j},\hp_{i]}],
	\end{align}	
\end{subequations}
such that noncommutativity of the coordinates $\theta^{ij}$ as well as the noncommutativity of the momenta $\tilde{\theta}_{ij}$ are entirely contingent on $f^i_j.$ 

According to the usual approach followed in the context of the GUP/EUP, the Hamiltonian of the system in question has to be expressed in terms of physical variables, \ie for a single particle of mass $m$
\begin{equation}
	H=\frac{\hp^2}{2m}+V(\hx^i,\delta_{ij}),
\end{equation}
with the flat metric $\delta_{ij},$ the squared physical momentum operator $\hp^2=\hp_i\hp_j\delta^{ij}$ and the potential $V.$ In the case of an harmonic oscillator, for example, the potential can be chosen to be of the form
\begin{equation}
	V_{\text{HO}}=\frac{m\omega \hbar^2}{2}\hx^2,
\end{equation}
such that it measures the squared distance from the origin.

An often pursued way to make sense of modified algebras of the kind \eqref{GUPalg} consists in finding a set of canonical phase space variables \cite{Bosso18b}, which the physical ones can be comprised of, and further constructing a Hilbert space within which the resulting Hamiltonian is Hermitian. This algorithm is equivalent to finding an explicit representation of the algebra. Along these lines, we define the new phase space coordinates noncanonically related to the physical ones $\hX=\hX(\hx,\hp),$ $\hP=\hP(\hx,\hp),$ which satisfy the Heisenberg algebra
\begin{align}
	[\hX^i,\hX^j]=[\hP_i,\hP_j]=0,\hspace{1cm}[\hX^i,\hP_j]=i\hbar\delta^i_j.\label{HeisAlg}
\end{align}
Then, the Hamiltonian can be expressed in terms of the said new variables as
\begin{equation}
	H=\frac{\hp^2(\hX,\hP)}{2m}+V[\hx^i(\hX,\hP),\delta_{ij}].
\end{equation}
As those can be represented in the usual way, the only ingredient left in the specific situation consists in finding the appropriate measure to make the resulting Hamiltonian Hermitian. As this can be quite demanding in the general case \eqref{GUPalg}, a formalism of the type introduced in \cite{Wagner21c} may be beneficial.

\section{Reinterpretation in terms of the curved cotangent bundle}\label{sec:reint}

It has recently been shown that the theory of a single particle obeying a general GUP can be mapped onto the dynamics of an ordinary particle moving on a curved momentum space \cite{Wagner21a,Wagner22}. The underlying derivation was based on the assertion that the correct extrapolation of the kinetic part of the Hamiltonian from flat to curved momentum space is of the form
\begin{equation}
	\hat{\tilde{H}}_{\text{kin}}=\frac{1}{2m}g^{ij}(\hP)\hP_i\hP_j,
\end{equation}
with the momentum-dependent metric $g^{ij}.$ However, this expression is not covariant in momentum space. In the truly reciprocal formulation pursued in the present work, covariance in position space has to be complemented by covariance in momentum space. Correspondingly, from the viewpoint of reciprocity and in line with the formalism of Relative Locality \cite{Amelino-Camelia11,Amelino-Camelia10}, it is more natural to define the free-particle Hamiltonian as
\begin{equation}
	\hat{H}_{\text{kin}}\equiv\frac{\hat{\sigma}_P^2(\hP)}{2m},\label{GeodDistHam}
\end{equation}
with the geodesic distance from the origin in momentum space $\hat{\sigma}_P.$ Its classical counterpart, defined as \cite{Synge60}
\begin{equation}
	\sigma_P\equiv\int_O^P\D s,\label{MomGeodDistDef}
\end{equation}
\ie the distance along the shortest geodesic (with $\D s$ being the corresponding line element) connecting the point in momentum space $P$ to the origin $O$, contains just as much information about the underlying manifold as the metric, which may, for example, be reconstructed from the differential version of Eq. \eqref{MomGeodDistDef}
\begin{equation}
	g_{ij}\frac{\partial \sigma_P}{\partial P_i}\frac{\partial \sigma_P}{\partial P_j}=1.\label{GeodDistDiff}
\end{equation}

Indeed, it has been recently pointed out that the dynamics derived from the classical limit of Hamiltonians of the form \eqref{GeodDistHam} yield the correct geodesics as solutions to the ensuing equations of motion even when both position and momentum space are curved \cite{Pfeifer21,Relancio21a,Relancio22} which rightly qualifies it as the kinetic energy.

A reinterpretation of the GUP in terms curvature in the cotangent bundle may therefore be based on the identification
\begin{equation}
	\hat{\sigma}_P^2=\delta^{ij}\hp_i(\hP )\hp_j(\hP ).
\end{equation}
When transitioning from the GUP to the GEUP, this relation can be kept as it is such that it becomes 
\begin{equation}
	\hat{\sigma}_P^2=\delta^{ij}\hp_i(\hX,\hP )\hp_j(\hX,\hP ).\label{IdentMom}
\end{equation}
However, it has to be complemented by an analogous construction in position space. Bearing in mind the principle of reciprocity, there is no freedom choice in that matter -- the missing piece must be the equality
\begin{equation}
	\hat{\sigma}_X^2=\delta_{ij}\hx^i(\hX,\hP)\hx^j(\hX,\hP),\label{IdentPos}
\end{equation}
with the geodesic distance operator in position space $\hat{\sigma}_X$, whose classical limit is defined analogously to Eq. \eqref{MomGeodDistDef}. By analogy with Eq. \eqref{GeodDistDiff}, this quantity can be related to the metric as
\begin{equation}
	g_{ij}\frac{\delta \sigma_X}{\delta X^i}\frac{\delta \sigma_X}{\delta X^j}=1,\label{PosGeodDistDiff}
\end{equation} 
where the positional derivative, which is modified by the curvature in the cotangent manifold, is defined as $\delta/\delta X^i=\partial/\partial X^i+N_{ij}\partial/\partial P_j,$ with the nonlinear connection $N_{ji}$ (for an exhaustive treatment of this subtlety consult \cite{Miron01,Miron12}).

Thus, the whole formalism is severely constrained by Eqs. \eqref{GUPalg}, \eqref{JacId}, \eqref{HeisAlg}, \eqref{IdentMom} and \eqref{IdentPos} and its consistency is not guaranteed \emph{a priori}. Therefore, it is instructive to investigate a prominent example.

\section{Perturbative quadratic GEUP}\label{sec:qGEUP}

From the phenomenological viewpoint, it is often sufficient to consider a perturbative version of the GEUP, which may be understood as an expansion in the origins of position as well as momentum space. In this context, momentum-dependent corrections to the Heisenberg algebra become important at high energies or small distances, \ie in the UV. This is why, they represent the minimal length paradigm, embodied by the GUP in the phenomenology of quantum gravity. Position-dependent modifications, in turn, gain relevance at large distances, \ie in the IR, the regime of classical General Relativity. In particular, they, summarized under the term EUP, are understood to model a maximal length implied by the existence of the cosmological horizon. 

\subsection{Algebra}

Usually, the combination of the said modifications can be parameterized with the help of the dimensionless coefficients $\alpha,$ $\alpha',$ $\beta$ and $\beta'$ as
\begin{align}
	[\hx^i,\hp_j]=&i\hbar \left[\delta^i_j\left(1+\alpha \hx^2/l_\text{IR}^2+\beta l_p^2\hp^2/\hbar^2\right)\right.\nonumber\\
	&\left.+\alpha'\hx^i\hx_j/l_\text{IR}^2+\beta'l_p^2\hp^i\hp_j/\hbar^2\right],
\end{align}
where $l_{\text{IR}}$ and $l_p$ denote a relevant length scale in the infrared (such as the radius of the horizon) and the Planck length, respectively. Correspondingly, the noncommutativities of the coordinates and momenta read approximately
\begin{align}
	[\hx^i,\hx^j]\simeq&i\hbar(2\beta-\beta')l_p^2\hat{j}^{ji}/\hbar^2\equiv i\hbar\theta \hat{j}^{ji},\\
	[\hp_i,\hp_j]\simeq&i\hbar(2\alpha-\alpha')\hat{j}_{ji}/l_\text{IR}^2\equiv i\hbar\tilde{\theta} \hat{j}^{ji},
\end{align}
with the modified angular momentum operator $\hat{j}_{ij}=2\hx_{[i}\hp_{j]}$ which, importantly, does not correspond to the generator of rotations (\cf \cite{Bosso16}) and where we introduced the noncommutativities in position and momentum space $\theta$ and $\tilde{\theta},$ respectively.  Given this algebra, we can study the equivalent representation in terms of the curved cotangent bundle.

\subsection{Transformation}

A general noncanonical transformation of interest to this work will be of the form $(\hx^i,\hp_j)\rightarrow (\hX^i,\hP_j)$
\begin{align}
	\hx^{ i}=&[1+c_\alpha\hX^2/l_\text{IR}^2+c'_{\beta}l_p^2\hP^2]\hX^i+c''_{\beta}l_p^2\hP^i\hP_jX^j,\\
	\hp_ i=&[1+c_\beta l_p^2\hP^2+c'_{\alpha}\hX^2/l_\text{IR}^2]\hX^i+c''_{\alpha}\hX_i\hX^jP_j/l_\text{IR}^2,\label{NonComTrafoPert}
\end{align}
where we abbreviated the quantities $\hX^2=\delta_{ij}\hX^i\hX^j$ and $\hP^2=\delta^{ij}\hP_i\hP_j$ and introduced the $\alpha$- and $\beta$-dependent dimensionless constants $c_\alpha,$ $c'_\alpha$,$c''_\alpha$, $c_\beta,$ $c'_\beta$ and $c''_\beta,$ respectively,

However, the formalism introduced in section \ref{sec:reint} puts severe constraints on the representable models. In particular, the covariance of the construction with respect to diffeomorphisms in position and momentum space requires that there is an infinite number of equivalently admissible transformations \eqref{NonComTrafoPert}, characterized by the gauge-parameters $g_\alpha$ and $g_\beta$. These label equivalent representations of the GEUP which are related by diffeomorphisms $(\hX^i,\hP_j)\rightarrow (\hX^{\prime i},\hP'_j)$ in position and momentum space 
\begin{align}
	\hX^{\prime i}=&[1+(g'_\alpha-g_\alpha)\hX^2/l_\text{IR}^2]\hX^i,&&\hP'_i=\frac{\partial \hX^j}{\partial\hX^{\prime i}}\hP_j,\\
	\hP'_i=&[1+(g'_\beta-g_\beta)l_p^2\hP^2]\hP_i,&&\hX^{\prime i}=\frac{\partial \hP_j}{\partial\hP'_i}\hX^j,
\end{align}
respectively.

Along the lines of section \ref{sec:reint}, Eqs. \eqref{GUPalg}, \eqref{JacId}, \eqref{HeisAlg}, \eqref{IdentMom} and \eqref{IdentPos} provide six conditions for the six constants numbers in Eq. \eqref{NonComTrafoPert}. However, the system of equation is only consistent if the gauge-parameters remain undetermined. Then, it does not come as a surprise that the model parameters have to be constrained as $\beta'=-\beta,$ $\alpha'=-\alpha$ instead. Thus, the algebra of the physical variables reads
\begin{align}
	[\hx^i,\hp_j]=&i\hbar\left[\delta^i_j+\beta l_p^2\left(\hp^2\delta^i_j-\hp^i\hp_j\right)/\hbar^2\right.\nonumber\\\
	&\left.+\alpha \left(\hx^2\delta^i_j-\hx^i\hx_j\right)/l_\text{IR}^2\right],\label{AllowedCom}
\end{align}
and the noncommutativities become $\theta=3\beta l_p^2/\hbar^2$ and $\tilde{\theta}=3\alpha/l_\text{IR}^2.$

This choice of parameters will be made for the remainder of this work. Although it has (to the knowledge of the author) not been considered yet in the literature, the model has some remarkable properties which will be dealt with below.

The solution of the system of equations reads in terms of the gauge-parameters
\begin{align}
	c_\alpha=&g_\alpha,&&c'_\alpha=\alpha-g_\alpha,&&c''_\alpha=-(2g_\alpha+\alpha),\\
	c_\beta=&g_\beta,&&c'_\beta=\beta-g_\beta,&&c''_\beta=-(2g_\beta +\beta).
\end{align}
Given the noncanonical transformation(s) and the constraints on model parameters, it is possible to analyse the dual theory on the curved cotangent bundle.

\subsection{Curvature}

In accordance with Eqs. \eqref{IdentMom} and \eqref{IdentPos}, the fully covariant harmonic potential energy and the kinetic energy operators are proportional to
\begin{align}
	\hat{\sigma}_x^2\simeq &\left[1+2g_\alpha\hX^2/l_\text{IR}^2+(\beta-g_\beta)l_p^2\hP^2/\hbar^2\right]\hX^2\nonumber\\
	&+(\beta-g_\beta)l_p^2\hX^i\hP^2\hX_i/\hbar^2\nonumber\\
	&-(2g_\beta+\beta)l_p^2\{\hX^i,\hP_i\hP_j\hX^j\}/\hbar^2,\\
	\hat{\sigma}_p^2\simeq &\left[1+2g_\beta l_p^2P^2/\hbar^2+(\alpha-g_\alpha)\hX^2/l_\text{IR}^2\right]\hP^2\nonumber\\
	&+(\alpha-g_\alpha)\hP_i\hX^2\hP^i/l_\text{IR}^2\nonumber\\
	&-(2g_\alpha+\alpha)\{\hP_i,\hX^i\hX^j\hP_j\}/l_\text{IR}^2.
\end{align}
Turning to commuting variables for the moment, the "classical" analogues of these quantities \footnote{The classical limit of the GUP is highly nontrivial, which can be understood heuristically through the disappearance of the length scale $l_p$ as $\hbar\rightarrow 0.$ For an in-depth analysis see \cite{Casadio20,Chashchina19}.}, the geodesic distances from the origin in position and momentum space, read 
\begin{align}
	\sigma_x^2\simeq &\left[1+2g_\alpha\hX^2/l_\text{IR}^2+2(\beta-g_\beta)l_p^2\hP^2/\hbar^2\right]X^2\nonumber\\
	&-(2g_\beta+\beta)l_p^2\left(P_iX^i\right)^2/\hbar^2,\\
	\sigma_p^2\simeq &\left[1+2g_\beta l_p^2P^2/\hbar^2+2(\alpha-g_\alpha)X^2/l_\text{IR}^2\right]P^2\nonumber\\
	&-(2g_\alpha+\alpha)\left(P_iX^i\right)^2/l_\text{IR}^2.
\end{align}
Using Eqs. \eqref{GeodDistDiff} and \eqref{PosGeodDistDiff}, the latter of which is simplified due to the nonlinear connection dropping out from the perturbative point of view, they can be understood as being derived from the metric
\begin{align}
	g_{ij}=&\left[1+2(\beta-g_\beta) l_p^2P^2/\hbar^2-2(\alpha-g_\alpha) X^2/\l_\text{IR}^2\right]\delta_{ij}\nonumber\\
	&-2(2g_\beta+\beta) l_p^2P_iP_j/\hbar^2\nonumber\\
	&+2(2g_\alpha+\alpha)X_iX_j/\l_\text{IR}^2.
\end{align}
Setting $\beta=\beta'=0$ corresponding to pure EUP-corrections, \ie a solely position-dependent metric as in usual curved spaces, the ensuing Riemann tensor $R_{ikjl}$ becomes
\begin{equation}
	R_{ikjl}\simeq \frac{6\alpha}{l_\text{IR}^2}\left(g_{ij}g_{kl}-g_{il}g_{kj}\right)\propto 2\tilde{\theta}.\label{Riemann}
\end{equation}
The case $\alpha=\alpha'=0,$ the GUP, can be treated by complete analogy with the preceding example (for a more in-depth treatment of this matter see \cite{Wagner21a,Wagner22}), thus implying that the curvature tensor in momentum space $S^{ikjl}$ equals
\begin{equation}
	S^{ikjl}\simeq \frac{6\beta l_p^2}{\hbar^2}\left(g^{ij}g^{kl}-g^{il}g^{kj}\right)\propto 2\theta.\label{Siemann}
\end{equation}
Both clearly indicate a space of constant curvature, which should be expected to lowest order in a newly introduced length scale. Furthermore, the parameters $g_\alpha$ and $g_\beta$ do not feature in the curvature tensors at all, corroborating the characterization of them being pure gauge.

In accordance with the results obtained in \cite{Wagner21a,Wagner22}, the scalar curvature in momentum space is proportional to the noncommutativity in of the coordinates in the original theory. An analogous relation is found for the curvature in position space and the noncommutativity of the momenta.

This does not come as a surprise because assuming $\beta=-\beta' $ (as well as $\alpha=0$ here) the representation in \cite{Wagner21a,Wagner22} is just one coordinatization of the results presented here. Moreover, the Born reciprocity of quantum mechanics dictates that modifications in position space behave analogously to those in momentum space.  It remains to be analysed which consequences this particular choice of model parameters entails. 

\subsection{Peculiarities of the model}

In order to translate the GEUP to the language of the curved cotangent manifold in a covariant way, we had to impose the constraint $\beta=-\beta'.$ This choice has some remarkable properties which are summarized in the present section.
\begin{itemize}
	\item[\textbullet] In one dimension, the commutator \eqref{AllowedCom} reduces to the canonical one, \ie the model is indistinguishable from quantum mechanics. This squares well with the fact that one-dimensional spaces cannot exhibit intrinsic curvature, while the model at hand necessarily implies a curved momentum space if $\beta\neq 0.$
	\item[\textbullet] Going hand in hand with the trivial one-dimensional limit, the usual limitations on length measurements are more subtle in the given model. For example, along the $i$th axis and for $\alpha=0$ and $\beta >0$ (\ie the usual GUP), we obtain
	\begin{equation}
		\Delta x^{(i)}\Delta p_{(i)}\geq \frac{\hbar}{2}\left(1+\frac{\beta l_p^2}{\hbar^2}\sum_{j\neq i}\Delta p_j^2\right),
	\end{equation}
	where the parentheses indicate indices which are not summed over. Clearly, the right-hand side of the uncertainty relation does not harbour any explicit dependence $\Delta p_N,$ but the momentum uncertainties along other directions.  In general, it is possible to imagine arbitrarily squeezed states such that $\Delta p_{N'}\neq \Delta p_{N'}(\Delta p_{N}).$ Then, the correction to the uncertainty relation amounts to a simple rescaling of Planck's constant $\hbar .$ Thus, there is no minimal length \emph{per se}; rather, the quantumness of phenomena along one axis depends on the resolution of momenta along the other ones.
	\item[\textbullet] As the coordinates continue being noncommutative, they still satisfy an uncertainty relation of the form
	\begin{equation}
		\Delta x^i \Delta x^j\geq \frac{3\beta l_p^2}{2\hbar}\braket{\hat{j}^{ji}}.
	\end{equation}
	This implies a minimal area, provided the system under study has nonvanishing angular momentum.
\end{itemize}
In short, the fate of the minimal length (and by analogy for $\alpha\neq 0$ the minimal momentum) is unclear and merits further study. 

To summarize, in the present section we have found the cotangent manifold geometry given a perturbative GEUP. Importantly, however, this construction not only applies one way: Given a geometry, we can find the dual description in terms of a GEUP. This is the matter of the next section. 

\section{Reverse engineering GEUPs\label{sec:revGEUPs}}

Curiously, the results \eqref{Riemann} and \eqref{Siemann} indicate that the modified commutator between positions and momenta \eqref{AllowedCom} underlying the GUP ($\alpha =0$) may be written as 
\begin{equation}
	[\hx^i,\hp_j]\simeq  i\hbar\left(\delta^i_j+S_j^{~kil}p_kp_l/6\right).
\end{equation}
Similarly, according to Eq. \eqref{Siemann} the noncommutativity of the coordinates is dependent on the curvature in the dual theory. An analogous relation can be written down on the EUP-side ($\beta=0$). 

This begs the question whether the form of the commutators is of general nature. In order to check this idea, we follow the approach presented in section \ref{sec:reint} inversely.  Afterwards, we apply the result to relativistic spacetime, in particular (anti-)de Sitter geometry.

\subsection{Ansatz}

Assume, for the moment, the background space to be described by the metric $g_{ij}(p),$ given in normal coordinates, and its associated Levi-Civita connection. Then, the geodesic distance from the origin (in momentum space) is given as $\hat{\sigma}_p^2=\hP^2,$ while in position space we obtain 
\begin{equation}
	\hat{\sigma}_x^2=\delta_{ij}e^i_k(\hP)\hX^ke^j_l(\hP)\hX^l,
\end{equation} 
with the \emph{vielbein} $e^i_j(\hP)=e^i_j(\hp).$ Thus, the noncanonical transformation towards the GUP-model can be  found nonperturbatively, reading simply 
\begin{equation}
	\hx^i =e^i_j\hX^i,\hspace{1cm}\hp_i=\hP_i.\label{NormalTrans}
\end{equation}
As a result, we obtain the modified algebra
\begin{align}
	[\hx^i,\hp_j]=&i\hbar e^i_j(\hp),\\
	[\hx^i,\hx^j]=&2i\hbar e^{[i}_k\mpar^{|k|}e^{j]}_l(e^{-1})^l_m(\hp)\hx^m,\\
	[\hp_i,\hp_j]=&0,
\end{align}
with the partial derivative in momentum space $\mpar^i=\partial/\partial \hp_i.$This is a completely general result, however \emph{only} applying to normal coordinates.

In order to obtain an explicit results, we expand the background in terms of Riemann normal coordinates centered in the origin on momentum space, such that
\begin{equation}
	e^i_j\simeq\delta^i_j+\left.S_j^{~kil}\right|_{p=0}\hp_k\hp_l/6,
\end{equation}
where the correction is positive because the metric in momentum space is the inverse of the metric in position space. Thus, the algebra can be approximated as
\begin{align}
	[\hx^i,\hp_j]\simeq & i\hbar\left(\delta^i_j+\left.S_j^{~kil}\right|_{p=0}\hp_k\hp_l/6\right),\\
	[\hx^i,\hx^j]\simeq & i\hbar\left.\left(S^{jikl}+S^{k[ij]l}\right)\right|_{p=0}\hat{j}_{kl}/6,\\
	[\hp_i,\hp_j]=&0,
\end{align}
where we used the symmetries of the Riemann tensor. Indeed, for a curvature tensor of the form \eqref{Siemann}, we obtain the GUP-side of the algebra \eqref{AllowedCom}. However, the GUP implied here is of more general scope, encompassing also anisotropic models (for more about these see \cite{Gomes22a,Gomes22b}).

Similarly, the quantum dynamics of a particle moving on a curved position space can alternatively be described by the EUP-deformed algebra
\begin{subequations}
	\label{CurvEUP}
	\begin{align}
		[\hx^i,\hp_j]\simeq & i\hbar\left(\delta^i_j+\left.R_{~kjl}^{i}\right|_{x=0}\hx^k\hx^l/6\right),\\
		[\hx^i,\hx^j]=&0,\\
		[\hp_i,\hp_j]\simeq & i\hbar\left.\left(R_{ijkl}+R_{k[ij]l}\right)\right|_{x=0}\hat{j}^{kl}/6.
	\end{align}
\end{subequations}
By analogy with the GUP, on a background curved as in Eq. \eqref{Riemann}, we exactly recover the algebra including \eqref{AllowedCom} for $\beta=0.$ Thus, we are in the position to provide a full derivation of EUPs under the assumption of a curved four-dimensional background.

\subsection{General EUP from spacetime curvature}

Ultimately, uncertainty relations in nonrelativistic single particle quantum mechanics have to be derivable from relativistic models. In this section, we show that a general EUP can be derived given an arbitrary spacetime $\mathcal{M}$ and a geometry on top.

The approach developed for nonrelativistic GEUPs in the preceding section can equivalently be applied to relativistic algebras. However, in this case its interpretability in terms of uncertainty relations is questionable; there are no relativistic uncertainty relations, even though there is no scarcity in ansätze \cite{Kudaka99,Todorinov:2018arx,Wagner21b}. The decisive reason behind this issue lies in there being no time operator in quantum mechanics, another problem that does not lack attempted solutions \cite{Susskind64,Kijowski:1974jx,Muga02}. Anyway, there is no harm in dealing with the underlying algebras on an abstract level.

In that vein, we can express the result of the preceding section in terms of Riemann normal coordinates in Minkowskian signature as
\begin{subequations}
	\label{CurvEUPMink}
	\begin{align}
		[\hx^\mu,\hp_\nu]\simeq & i\hbar\left(\delta^\mu_\nu+\left.R_{~\rho\nu\sigma}^{\mu}\right|_{x=0}\hx^\rho\hx^\sigma/6\right),\\
		[\hx^\mu,\hx^\nu]=&0,\\
		[\hp_\mu,\hp_\nu]\simeq & i\hbar\left.\left(R_{\mu\nu\rho\sigma}+R_{\rho[\mu\nu]\sigma}\right)\right|_{x=0}\hat{j}^{\rho\sigma}/6.
	\end{align}
\end{subequations}
Given the abstractness of this expression, it is instructive to demonstrate its consequences in terms of a simple explicit example.

\subsubsection*{(Anti)-de Sitter space}

As (anti)-de Sitter space is maximally symmetric, in the origin of Riemann normal coordinates the Riemannian curvature tensor can be expressed as
\begin{equation}
	R_{\mu \rho\nu\sigma}|_{x^\mu=0}=\frac{\Lambda}{3}\left(\delta^{\mu\nu}\delta^{\rho\sigma}-\delta^{\mu\sigma}\delta^{\rho\nu}\right),
\end{equation}
with the cosmological constant $\Lambda$ which is (negative) positive for (anti-) de Sitter space. As a result, we obtain the algebra
\begin{subequations}
	\label{CurvEUPAdS}
	\begin{align}
		[\hx^\mu,\hp_\nu]\simeq & i\hbar\left[\delta^\mu_\nu \left(1+\frac{\Lambda}{18}\hx^2\right)-\frac{\Lambda}{18}\hx^\mu\hx_\nu\right],\\
		[\hx^\mu,\hx^\nu]=&0,\\
		[\hp_\mu,\hp_\nu]\simeq & i\hbar\Lambda\hat{j}_{\nu\mu}/6.
	\end{align}
\end{subequations}
Note that this algebra is still coordinate dependent, while the Hamiltonian is coordinate invariant. For example, we may apply the coordinate rescaling $\hx^\nu\rightarrow \sqrt{2}\hx^\mu,$ $\hp_\mu\rightarrow\hp_\mu/\sqrt{2}$ to bring the commutation relations more in line with the usual way they are presented such that
\begin{subequations}
	\label{CurvEUPAdS2}
	\begin{align}
		[\hx^\mu,\hp_\nu]\simeq & i\hbar\left[\delta^\mu_\nu \left(1+\frac{\Lambda}{9}\hx^2\right)-\frac{\Lambda}{9}\hx^\mu\hx_\nu\right],\\
		[\hx^\mu,\hx^\nu]=&0,\\
		[\hp_\mu,\hp_\nu]\simeq & i\hbar\Lambda\hat{j}_{\nu\mu}/3.
	\end{align}
\end{subequations}
Clearly, this result is not exactly the position space analogue of Snyder-space, which, however, is not required for consistency. 

\section{Discussion}\label{sec:Discussion}

Modified Heisenberg algebras play a significant r\^{o}le in the modern discussion on quantum gravity phenomenology. In this paper we have presented a map that provides an alternative description of a nonrelativistic quantum particle obeying a quadratic GEUP, which contains position- and momentum-dependent corrections embodying classical (maximal length) and quantum (minimal length) gravity effects, in terms of the quantum dynamics of an ordinary particle on a generally curved phase space. In so doing, we have consistently respected the Born reciprocal symmetry of the undeformed quantum theory implying results which are covariant in position as well as momentum space. This has been achieved by reinterpreting the kinetic term of the Hamiltonian and the harmonic oscillator potential as proportional to the squared geodesic distances to the origins in momentum and position space, respectively.

In comparison to earlier results \cite{Wagner21a,Wagner22}, the covariant nature of the results presented here leads to constraints on admissible GEUPs for description in terms of the curved cotangent manifold. Somewhat reminiscently of Loop Quantum Gravity and in contrast to conventional approaches, its GUP-sector does not exhibit a minimal length but noncommutative coordinates; as is to be expected of curvature effects, its consequences cannot be restricted to solely one dimension. Therefore, this model arguably merits further consideration in the future in its own right.

We have further found that, on the one hand, the case of pure EUP-corrections results in a solely position-dependent metric and a Riemann tensor that is proportional to the noncommutativity of the momenta in the original theory. On the other hand, the pure minimal-length limit, implying a GUP, yields the same behaviour for the curvature tensor in momentum space and the noncommutativity of the coordinates as expected by Born reciprocity.  

As a further application, we have reversed the formalism to derive GUPs and EUPs from general geometries on curved phase space. Thus, we have expressed the widely considered isotropic as well as the more general anisotropic models in terms of curvature tensors. Moreover, we have considered relativistic curved position spaces and their corresponding relativistic EUPs by analogy. To provide an example, we have applied the formalism to (anti-)de Sitter spacetimes.

Note that all derivations in the present work have been performed at first perturbative order. Whether this behaviour persists to higher order in the perturbative expansion will be the subject of future work. 

\section*{Acknowledgments}
	
	The author thanks P. Bosso and L. Petruzziello for insightful discussions. His work was supported by the Polish National Research and Development Center (NCBR) project ''UNIWERSYTET 2.0. --  STREFA KARIERY'', POWR.03.05.00-00-Z064/17-00 (2018-2022). Moreover, he would like to acknowledge the contribution of the COST Action CA18108.
	
\bibliographystyle{apsrev4-1}
\bibliography{ref}
	
\end{document}